# Using a market basket analysis in tourism studies

Damjan Vavpotič*
Faculty of Computer and Information Science, University of Ljubljana,
Večna pot 113,1000 Ljubljana, Slovenia
P: +386 1 479 8228; e-mail: damjan.vavpotic@fri.uni-lj.si

Karmen Knavs
Faculty of Computer and Information Science, University of Ljubljana,
Večna pot 113, 1000 Ljubljana, Slovenia
e-mail: karmen.knavs@gmail.com

Ljubica Knežević Cvelbar
Faculty of Economics, University of Ljubljana,
Kardeljeva Ploščad 17, 1000 Ljubljana, Slovenia
P: +386 1 5892 400; e-mail: ljubica.knezevic@ef.uni-lj.si

*corresponding author

**Original contribution available in Tourism Economics:**

**Vavpotič D, Knavs K, Knežević Cvelbar L, Using a market basket analysis in tourism studies, Tourism Economics (First Published July 30, 2020), © 2020 (SAGE Publications). DOI: 10.1177/1354816620944264.**

**Abstract**

Understanding tourist visitation patterns is crucial for decision makers in order to create smart tourism industry. A growing body of tourism research uses geo-location data in order to better understand tourism demand. In this paper, we present a new approach based on a market basket analysis. This approach uses geo-location data shared by tourists on tourism platforms in order to bundle the range of available tourism services and understand which experiences are consumed together. The approach was tested on the case of Vienna, Austria. Based on our analyses we argue that the proposed approach has potential for use at the destination level and provides relevant information on tourism demand patterns important for smart tourism decision-making.

## Introduction

There have been a growing number of papers using user-generated content data, specifically geo-location in order to identify tourism flows in a certain region (Grinberger et al., 2014; Jin et al., 2018; Li and Yang, 2017; Önder et al., 2016). Tourism flows are defined as repetitive movements of tourists within a certain geographical area (Reinhold et al., 2015). Data were so far used from Flickr™, Twitter™, Instagram™, TripAdvisor™ and other social networks. Lately we have also observed examples of using mobile phone signal data to track the movement of tourists within a space (Raun et al., 2016). Those data are much more accurate and detailed and therefore provide a great potential for a better understanding of tourism flows.

Most of the studies (Shoval, 2012; Shoval and Ahas, 2016; Yang et al., 2013) use user- or automatically-generated data to explore tourism flows in a certain geographical area. Tourism flows are then identified for broader tourism regions and they provide information that is of great use for tourism destination managers and operators. Based on the identified tourism flows, destination managers and operators can better plan their marketing and management activities.

Using the tourism flow logic at the level of a single tourist destination has certain limitations too. Destinations can be quite different in terms of scale or products provided and are located within a limited geographic space. Capturing the sequence of tourism movements within such a limited geographic space where the attractions are only a few hundred meters apart and/or quickly reachable by public transport, is less relevant. In such settings, we can identify many different sequences that are mostly a consequence of casual tourist decisions made on the spot without a single dominant sequence. Therefore, our study focuses on the question of which attractions are visited together.

We are proposing market basket approach to help us better understand tourist visitation patterns at the destination level. A market basket is a set of products that are bought or consumed at the same time. This logic is commonly used in retailing in order to better understand human behaviour. We applied this logic to tourism in order to determine the set of tourist experiences that are consumed by tourists during their visit to a certain tourist destination. Paper is therefore partially tacking field of behavioural economics since it enables to better understand the behavioural patters of tourists. To our knowledge, this methodological approach has not been used in tourism industry to date. The empirical case used in this paper is Vienna, Austria. We collected user-generated data, 212,414 posts shared by tourists on TripAdvisor™ and used this data to create attraction sets, build co-occurrence graphs and preform network community detection analysis using Infomap™ algorithm. The results of our analysis are crucial for understanding tourist visitation patterns in tourism at the level of tourist destinations.

## Web and automatically-generated data used to define tourism flows

Digitalization of society can in many ways help destinations to better manage tourism demand. Large amounts of data are collected automatically and can already today be monitored in real time (for instance, Google Maps™ are offering real-time data to monitor and inform us about traffic). There are also large amounts of data being shared by users of various platforms that can be useful to understanding tourism demand better. With the availability of such big data new avenues in tourism research are now opened.

These studies mainly used geo-location for understanding the spatial patterns and movements of tourists and project tourism-flows within the certain geographical areas.

A good example are advanced studies based on the analysis of data collected through the global positioning system (GPS). Those data use location information in order to project the movement of tourists in a certain geographic area. One of the first studies based on GPS location information was conducted by Shoval (2008). In this study tourists were carrying GPS devices that collected their location information. Unfortunately, due to technical issues only a bit more than a half of the data could be collected. Yet the collected data allowed the author to reconstruct the tourists' movement and better understand their visitation patterns and movements. Those data provided migration patterns of tourists within the space, but did not provide insights in tourists' characteristics. Authors are now pursuing further development of this approach by using GPS data along with survey studies (Bauder and Freytag, 2015; De Cantis et al., 2016; Pettersson and Zillinger, 2011), interviews (Edwards and Griffin, 2013) and SMS services in which tourists can define their characteristics in more detail (Birenboim et al., 2015; Reinau et al., 2015).

Studies using actual GPS devices were abolished with the availability of user-generated data that can capture publicly shared opinions, reviews, photos, etc. (Grinberger et al., 2014) While reviewing a hotel or attraction, tourists also share their travel experience through photos or videos. This information typically contains time and geographical location that provides an understanding of the migration of tourists and their visitation within a certain space.

One of the first studies that used user-generated data to project tourism flows was published by Girardin, Fiore, Ratti, and Blat (2008) in *Journal of Location Based Services*. The authors used geographically referenced photos taken by 4,280 photographers over a period of two years in the Province of Florence, Italy. Based on those data, they created visualizations of tourist concentration and spatial-temporal flows. The paper was published in a non-tourism journal (*Journal of Location Based Services*), which is why it did not immediately attract the attention of the tourism academic community.

More studies using geo-location and time data to define tourism flows have been published since 2014. Kádár (2014) identified spatial patterns of tourist activities between different European cities based on registered bed nights and geotagged tourist photos shared on Flickr™. Vu et al. (2015) used 29,443 photos posted on Flickr™ collected from 2,100 inbound tourists visiting Hong Kong. Based on spatial and temporal information data they reconstructed tourists’ movements within the space. Furthermore, Amaral et al. (2014) used data shared on TripAdvisor™ to evaluate tourist experiences with restaurant services and define groups of tourists visiting similar restaurants.

García-Palomares et al. (2015) used photos from eight major European cities shared on Panoramio™. They distinguish between photos posted by locals and by tourists. Their spatial distribution patterns were analysed using spatial statistical techniques in a geographic information system (GIS). The result has shown significant differences in spatial distribution patterns between locals and tourists, where tourists show higher spatial concentrations. Chua et al. (2016) identify the tourism flows in Cilento – a regional tourist attraction in southern Italy – based on geotagged social media data from Twitter™. Data on spatial, temporal and demographic characteristics of tourists are used to identify tourism flows. Hauthal and Burghardt (2016) extracted location-based emotions from geo-referenced photos of Dresden published on Flickr™ and Panoramio™, and constructed an emotional map to provide insight into tourist visitation. Authors used emotions that describe the places such are: boring, attractive, scary, exciting etc. and photos published along to those emotional statements to construct the emotional map.

Shi et al. (2017) explored urban tourism crowding in Shanghai using crowdsourcing geospatial data from Weibo™. Two studies used user-generated content to visualize tourism flows of event visitors. Leung et al. (2012) used online trip diaries to analyse overseas tourist movement patterns during the Beijing Olympics. Kirilenko and Stepchenkova (2017) analysed Twiter™ comments to define visitor movements during the Sochi Winter Olympics.

Salas-Olmedo et al. (2018) used data from multiple user-generated sources (Panoramio™, Foursquare™ and Twitter™) to identify tourism flows. They concluded that several web-generated sources of data are providing a better insight in tourists' spatial movements than a single data source. Jin et al. (2018) used user generated content (UGC) collected from an open tourism web service to identify tourism flows in Nanjing City. Cvelbar et al. (2018) used data shared on TripAdvisor™ to identify visitor flows in the North East Adriatic Region. They identified 31 visitor flows between 188 destinations in the region.

Ahas et al. (2008) used passive mobile positioning data for the first time, which is automatically stored by mobile operators, and analysed them on the case of Estonia. The results indicated that passive mobile positioning data are in line with conventional accommodation statistics. In this study, data were not analysed in depth and tourism flows were not identified. The study opened a discussion about the potential for advanced data analytics using mobile positioning data. The next study on the empirical case of Estonia was publishes by Raun et al. (2016) and they used mobile phone positioning data of foreign visitors in Estonia from 2011 to 2013 to identify visitor flows. Using a mobile phone positioning data significantly improved data quality and accuracy and therefore also the reliability of results.

**Market basket analysis**

Market basket analysis (MBA) in an important datamining technique initially developed to help businesses identify purchasing behaviour. There is an extensive number of studies using MBA in retail settings where its results may be used to improve supplier's profitability, quality of service and customer satisfaction (Kutuzova and Melnik, 2018). In MBA Apriori algorithm (Agrawal and Srikant, 1994) and its derivatives are often used to determine frequent item sets and association rules (Chen et al., 2005; Cheng et al., 2015; Videla-Cavieres and Ríos, 2014). However, due to Apriori algorithm limitations alternative algorithms were developed to perform MBA. These algorithms rely on network analysis and graph mining techniques to detect network communities (Faridizadeh et al., 2018; Videla-Cavieres and Ríos, 2014).

MBA assumes that more than one item from a number of product categories are included in a single purchase (Russell and Petersen, 2000) which is typical for retail shopping. However, this is also likely in the context of tourism purchasing decisions, where a tourist has to purchase transport, accommodation, entertainment, food, etc. as a part of a single travel (Solnet et al., 2016). Although use of MBA and network analysis approaches in the context of tourism is limited (Tran et al., 2016), we can expect its popularity in tourism to increase along with other big data-related research (Li and Law, 2020).

## The proposed approach

The proposed approach used in this paper, based on tourist geo-location data, builds on existing approaches for market basket analysis (MBA) in retail (Chen et al., 2005; Videla-Cavieres and Ríos, 2014). The idea is to assign visits of attractions made by a certain tourist in a limited time period and a limited geographical area to a common set. A similar approach is used in MBA to create a set of products purchased by a certain customer in retail merchandising scenarios (Videla-Cavieres and Ríos, 2014).

However, in retailing one can rely on receipts that accurately record the products purchased together. Furthermore, it is more difficult to obtain data about visited attractions, as it is impossible to rely on receipts. Specifically, many attractions (such as public places in a city, natural wonders, etc.) are free of charge and even if they do charge a fee, it is often impossible to attribute purchases of different attractions to a certain tourist as purchases are paid anonymously in cash or group purchases are made. To overcome this obstacle, we use publicly shared posts created by tourists on the tourism web platform TripAdvisor™. We selected TripAdvisor™ as the largest reliable source of data about attractions. Tourist posts on TripAdvisor™ are organized, clearly assigned to attractions and monitored by TripAdvisor™. This is not the case with other data sources such as Instagram™, Flicker™, TikTok™ etc.

Each TripAdvisor™ post contains three key values used in our analysis: the identifier of a visited attraction, the (anonymous) identifier of a tourist who submitted the post, and the date of post submission. The proposed approach shown in Fig. 1 consists of four phases: 1) creation of attraction sets, 2) creation of a co-occurrence graph, 3) network community detection and 4) visualisation of community networks.

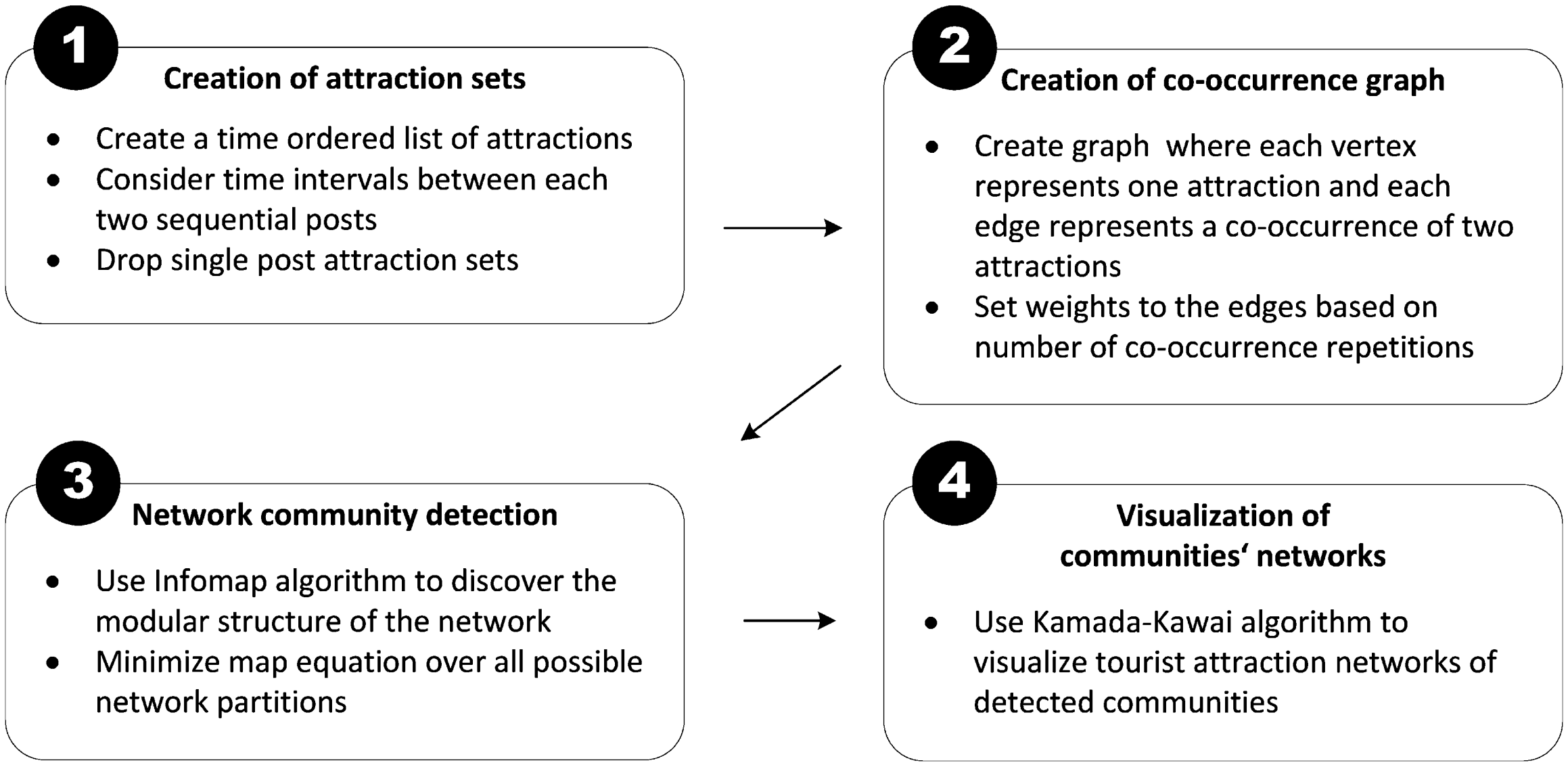

*Figure 1: The proposed approach.*

In the first phase, attraction sets are created in the following three steps: In the first step, we use the identifier of the tourist, the identifier of the attraction and the date of post submission to create a time-ordered list of attractions visited by a certain tourist in a selected geographical area. In the second step, we focus on time intervals between each two sequential posts from the time ordered list of attractions. Short time intervals (e.g. a day, several days, a week) between any two sequential posts created by the same tourist indicate a high probability that the two attractions were visited during the same visit and should therefore be assigned to a common attraction set. However, long time intervals (e.g. several weeks, a month, a year) between two sequential posts indicate two separate visits that have to be assigned to two separate attraction sets. We start the creation of attraction sets by computing the length of the time interval between the first and the second post in the sorted list. If the length of the interval between the two posts is shorter or equal to a certain predefined value, we create one attraction set for both posts, otherwise we create two attraction sets, one for the first and one for the second post. We continue by computing time intervals between the second and the third post and assign the third post either to the same attraction set as the second post, or to a new separate attraction set. This process is repeated until each post from the sorted list is assigned to an attraction set. In the third (i.e. last) step of the first phase, all attraction sets consisting of only a single post are dropped as they do no indicate any co-occurrence (just a single post) and therefore cannot be used in further co-occurrence analysis.

In the second phase, the attractions sets are used to create a co-occurrence graph where each node represents one attraction and each edge represents a co-occurrence of two attractions in attraction sets. Based on the number of co-occurrence repetitions of attractions in attraction sets we assign weights to the edges. The weights denote how many times two specific attractions were visited together by different tourists. For instance, if only one tourist visited the two specific attractions together the weight would be one, if two different tourists visited the two specific attractions together the weight would be two, etc. The resulting co-occurrence graph is actually an undirected weighted network of attractions. However, one major drawback of networks is that, for analytical and visualization purposes, they can only depict small systems (Rosvall et al., 2009). The interpretational value of visualisation of larger networks is limited due to a number of overlapping nodes and connections. A possible solution to this issue is to focus on communities, i.e. the subsets of nodes which are more densely connected internally than with other nodes within the network. In our case, focusing on communities is very convenient as they actually represent the subsets of paid and unpaid attractions that are frequently visited together.

In the third phase, we detect communities of attractions. The purpose of community detection is to discover the modular structure of the network. Different algorithms have been developed for detection of communities in networks (Agreste et al., 2017). In our case, an algorithm was required that would identify communities on a relatively large network with undirected weighted links. Based on the existing comparison of suitability of different approaches (Agreste et al., 2017), we selected Infomap™ algorithm (Bohlin et al., 2014) as a suitable option. Infomap™ is a network clustering algorithm based on the map equation (Rosvall et al., 2009). It was successfully used in many community detection problems and provides the best trade-off between accuracy and computational costs (Agreste et al., 2017). It is based on the information-theoretic method map equation, which in contrast to modularity optimisation approaches (Blondel et al., 2008) attends to identify patterns of flow on the network (Rosvall et al., 2009). In our case, sequential visits of different attractions made by tourists can be considered “a flow”. This flow inherently becomes part of attraction sets created in the first phase of our approach and is then propagated to the co-occurrence network produced in the second phase. It is therefore important that the selected algorithm for the detection of communities, i.e. Infomap™, considers this dynamic nature of the co-occurrence network. Infomap™ algorithm relies on a random walker as a proxy for real flow in the network. A random walker is a stochastic process that crates a path consisting of a sequence of random steps in the network. If a network has strong community structure, one expects that a random walker within a community occupies a majority of steps if one optimally partitions the network into communities (Masuda et al., 2017). Infomap algorithm uses map equation as a measure of theoretical limit of how concisely the trajectory of a random walk can be described on a selected module partition M. By minimizing the map equation (L(M)) over all possible network partitions, we can identify important aspects of the network structure with respect to the dynamics on the network (Rosvall et al., 2009). Map equation for undirected weighted networks that we use in our case can be expressed as (Rosvall et al., 2009):

$$L(M) = w_{\curvearrowright} \log(w_{\curvearrowright}) - 2\sum_{i=1}^{m} w_{i\curvearrowright} \log(w_{i\curvearrowright}) - \sum_{\alpha=1}^{n} w_{\alpha} \log(w_{\alpha}) + \sum_{i=1}^{m} (w_{i\curvearrowright} + w_i) \log(w_{i\curvearrowright} + w_i)$$

where L(M) is map equation for module partition M of n nodes ($\alpha = 1, \dots, n$) into m modules ($i = 1, \dots, m$), $w_\alpha$ is the relative weight of node α where the relative weight is defined as the total weight of the links connected to the node divided by twice the total weight of all links in the network, $w_i$ is the relative weight of the module i defined as $w_i = \sum_{\alpha \in i} w_\alpha$, $w_{i\curvearrowright}$ is the relative weight of the links exiting module i, and $w_\curvearrowright$ is the total relative weight of links between modules defined as $w_\curvearrowright = \sum_{i=1}^{m} w_{i\curvearrowright}$.

If the fourth phase, we examine the modular structure of the network by visualising the key communities. Understanding the communities in the attraction network enables us to identify bundles of attractions visited together, determine the level of attractions' reciprocal influence in terms of generating tourism flow, understand the importance of different parts of the network, etc. Different approaches to the visualisation of an undirected network are possible. The one that we found very useful for visualising tourist attraction networks is the well-established Kamada-Kawai algorithm (Kamada and Kawai, 1989).

## Evaluation of the proposed approach

### Case description

We evaluated the proposed approach on the case of Vienna, the capital of Austria. We selected Vienna as it is a well-developed tourist destination that promotes various highly visited tourist attractions. The official statistics show 15.5 million tourist overnight stays in 2017 (Lukacsy and Fendt, 2018). Our main aim was to evaluate the feasibility and usefulness of the proposed approach.

We analysed 212,414 tourist posts related to 519 different Vienna attractions that were submitted to TripAdvisor™ from Jun 4$^{th}$ 2002 and until Mar 5$^{th}$ 2018. Fig. 2 shows the number of posts over years. The number of posts increased significantly in the last decade. For instance, in 2017 more than 47,000 Vienna-related posts were submitted, while a decade earlier in 2007 there were only 56 such posts.

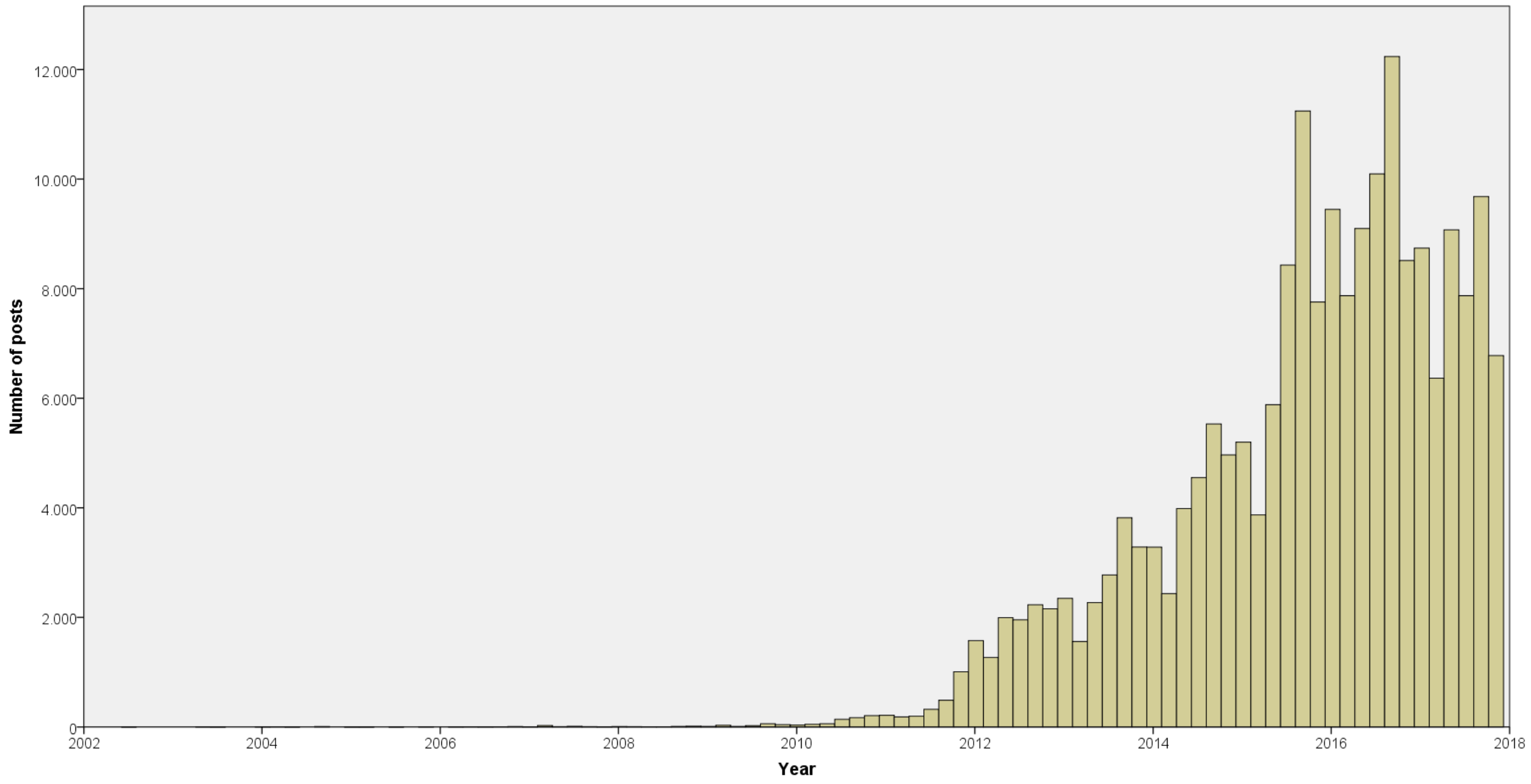

*Figure 2: Number of posts related to Vienna attractions between Jun 4th 2002 and Mar 5th 2018.*

In addition, the number of posts for different attractions varied considerably. The most visited attraction (Schoenbrunn Palace) had over 30,000 posts, while 210 attractions had 10 posts or less. Fig. 3 shows the number of tourist posts for the 519 attractions on a logarithmic scale.

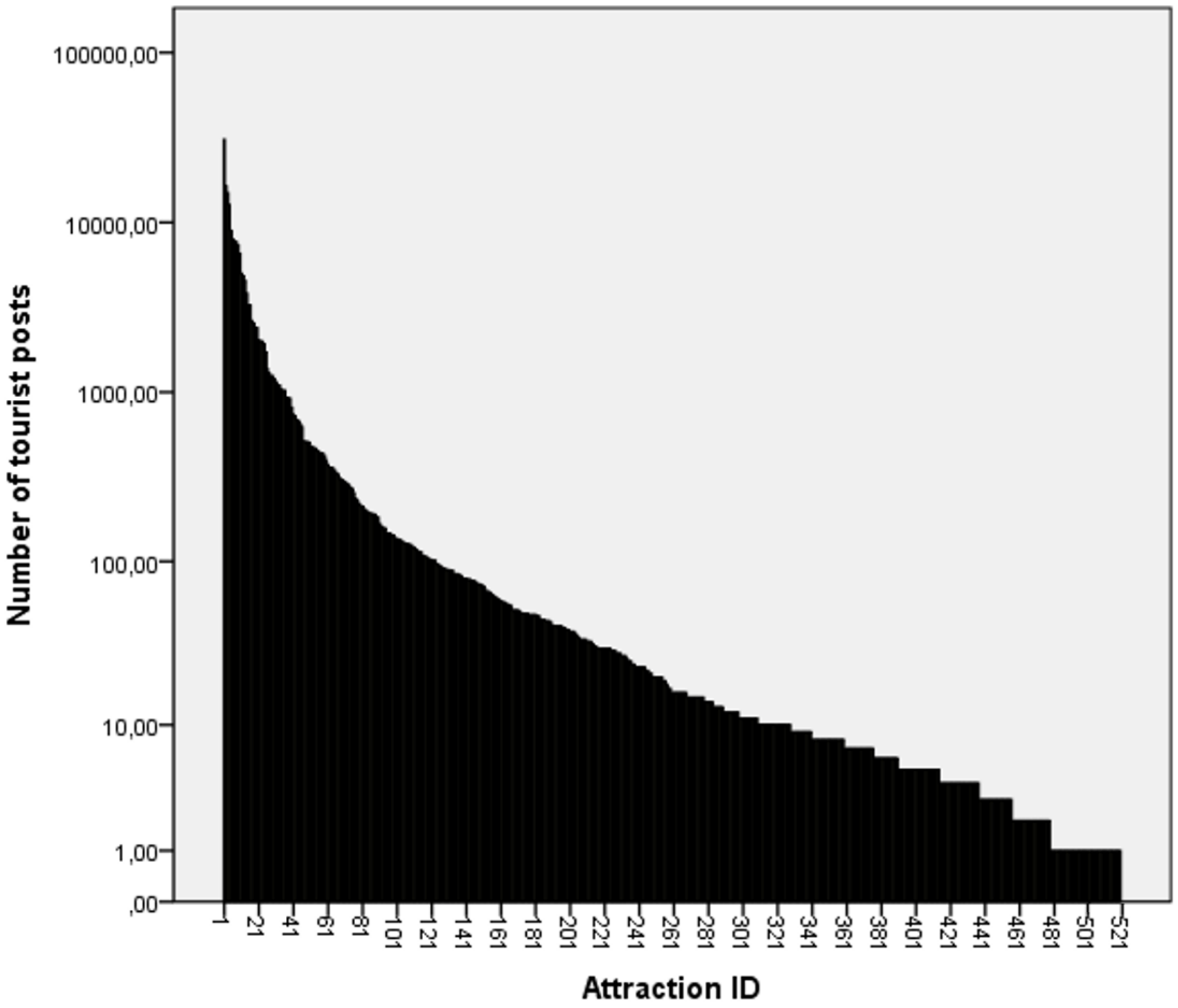

*Figure 3: Number of posts for the 519 Vienna attractions on a logarithmic scale.*

**Application of the proposed approach with the results and discussion**

In line with the proposed approach, we first used the posts to create attraction sets (first phase) and next used the attraction sets to create a co-occurrence graph (second phase). The obtained co-occurrence graph encompassed 40,503 undirected weighted edges where each edge represented a co-occurrence of two specific attractions and the weight of the edge represented the number of repetitions of this co-occurrence. Similarly, as in many other big data sets, most of the edges had very low weights, which means that only a few tourists reported visiting the two co-occurring attractions in close time proximity. Such edges are of little significance for model building and subsequent inferences and can be removed. However, there is no single correct method to select the minimum weight below which the edges can be removed (Kosinski et al., 2016). After experimenting with different minimum required weights, we retained the edges with weights higher than five i.e. co-occurrences with at least 5 repetitions. This threshold enabled us to remove 31,574 poorly represented co-occurrences and reduce network complexity while still retaining a sufficient amount of information for the analysis.

The resulting co-occurrence graph comprised 255 attractions and 8,929 co-occurrences. We used the Infomap™ algorithm to detect communities in this graph (third phase). Infomap™ algorithm was set to overlapping mode that allows attractions to be part of more overlapping communities. Infomap™ algorithm detected six communities, of which the largest one was comprised of 250 attractions and 8,917 co-occurrences, while the remaining five comprised from two to five attractions and from one to four co-occurrences. The six communities and their connections are shown in Fig. 4 (communities are marked with letters from A to F). The largest community (A) contained over 99 % of all visitations in the network. This clearly indicates that tourism flows in Vienna are concentrated around one central community containing attractions that are frequently visited together. As presented in Fig. 4, we also have five smaller communities (marked with letters from B to F) connected only to this central community A. The size of a community indicates the relative volume of the network flow inside the community, while the length and the thickness of the connections indicate the relative volume of the network flow between two connected communities. Fig. 4 shows that the relative volume of the network flow between the communities F and A is the lowest while there is some more volume between C, E and A.

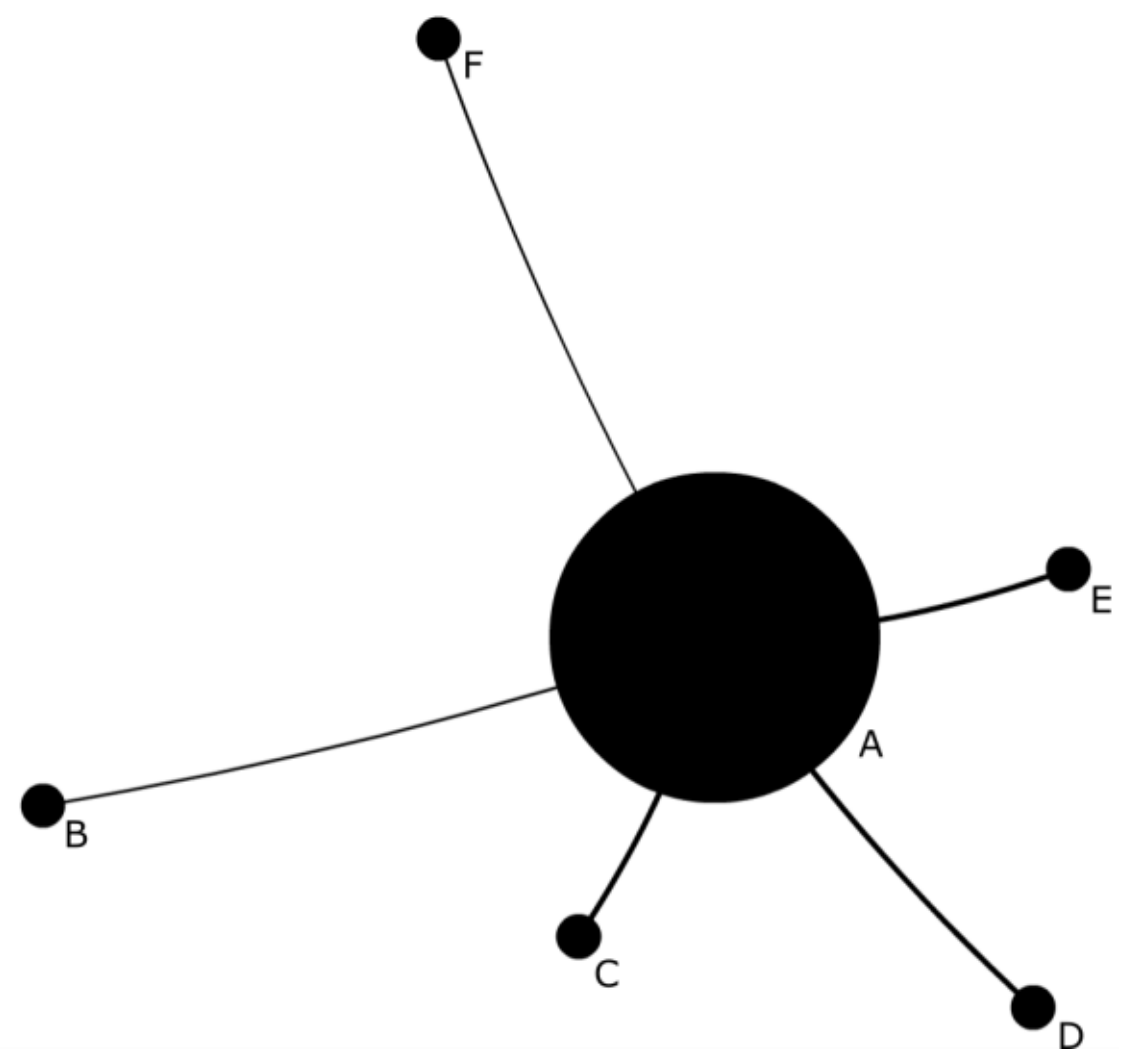

*Figure 4: The six detected communities (A–F) and their connections.*

The most interesting is the largest community A, which we describe in detail. The distribution of the network flow volume for the 250 attractions of community A having logarithmic nature is shown in Fig. 5 and the exact share of the network flow for the top 20 attractions is presented in Table 1.

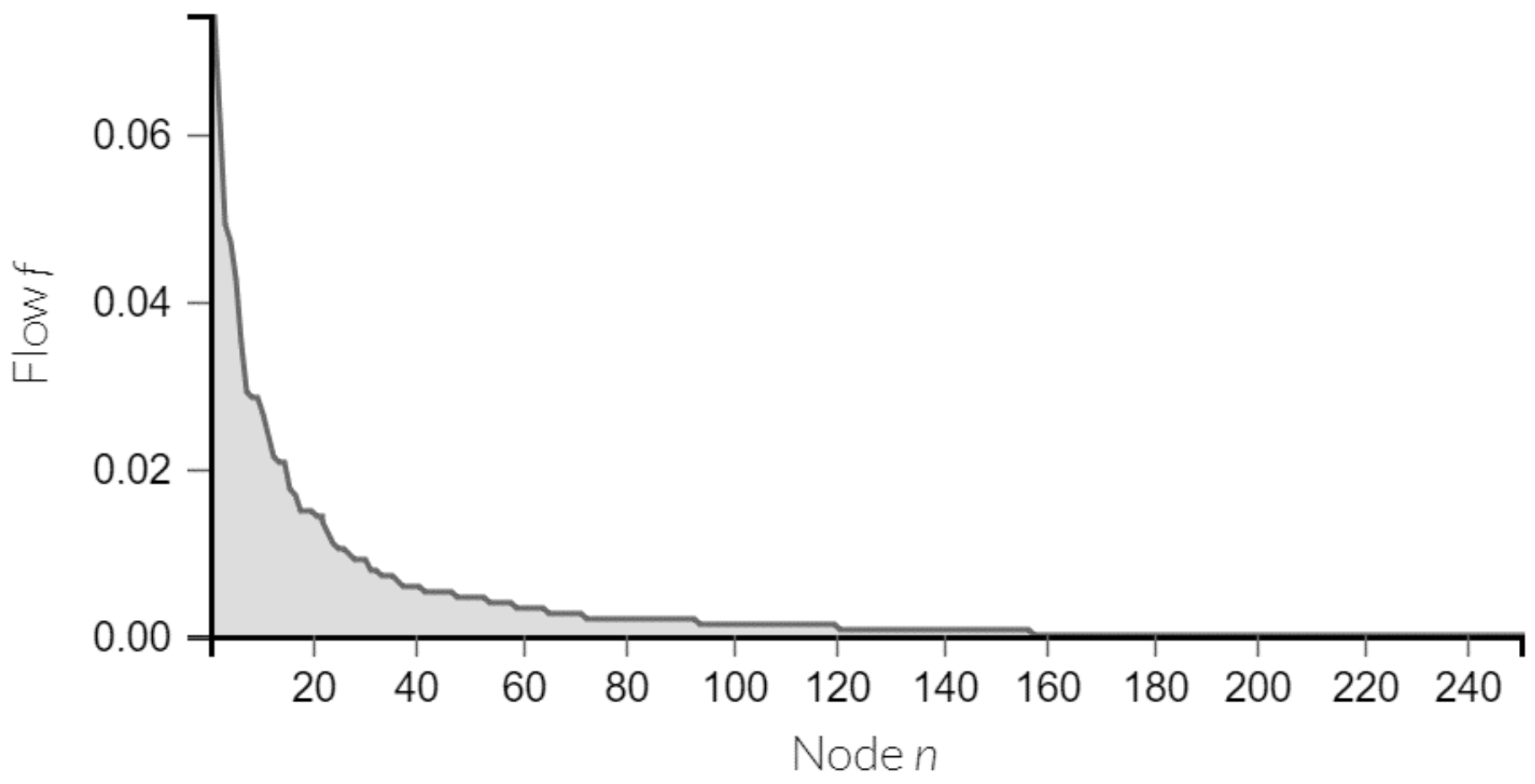

*Figure 5: Distribution of the network flow volume in community A.*

As we can see from Table 1 top 20 attractions contain over 60% of total network flow, while the top five attractions contain approximately 28 %.

*Table 1: Top 20 attractions (nodes) of the community A that together contain over 60 % of total network flow.*

| Nu. | Node (attraction) | % of the network flow |
|---|---|---|
| 1 | Schonbrunn Palace | 7.43 |
| 2 | St. Stephen's Cathedral | 6.20 |
| 3 | Historic Center of Vienna | 4.93 |
| 4 | Belvedere Palace Museum | 4.69 |
| 5 | Imperial Palace (Hofburg) | 4.27 |
| 6 | Schonbrunner Gardens | 3.46 |
| 7 | Stephansplatz | 2.91 |
| 8 | Prater | 2.85 |
| 9 | Kunsthistorisches Museum | 2.83 |
| 10 | Rathaus | 2.63 |
| 11 | State Opera House | 2.41 |
| 12 | St. Peter's Church | 2.10 |
| 13 | Tiergarten Schoenbrunn - Zoo Vienna | 2.07 |
| 14 | Albertina | 2.05 |
| 15 | MuseumsQuartier Wien | 1.77 |
| 16 | Ringstrasse | 1.68 |
| 17 | Natural History Museum | 1.50 |
| 18 | Sisi Museum | 1.47 |
| 19 | Karlskirche | 1.46 |
| 20 | Rathausplatz | 1.41 |

In Fig. 6 we visualize the top 45 attractions and 300 co-occurrences that are part of community A and contain approximately 80 % of the total network flow. The points represent attractions where the size of each point indicates the relative volume of the network flow through the attraction, while the length and the thickness of each connection indicates the relative volume of the network flow between the two connected attractions. The visualization clearly shows the importance of Vienna's main attractions: "Schonbrunn Palace", "St. Stephen's Cathedral", "Historic Center of Vienna", "Belvedere Palace Museum" and "Imperial palace (Hofburg)". These five attractions together account for approximately 28 % of the total network visitations (see Table 1) and are directly connected to many other attractions.

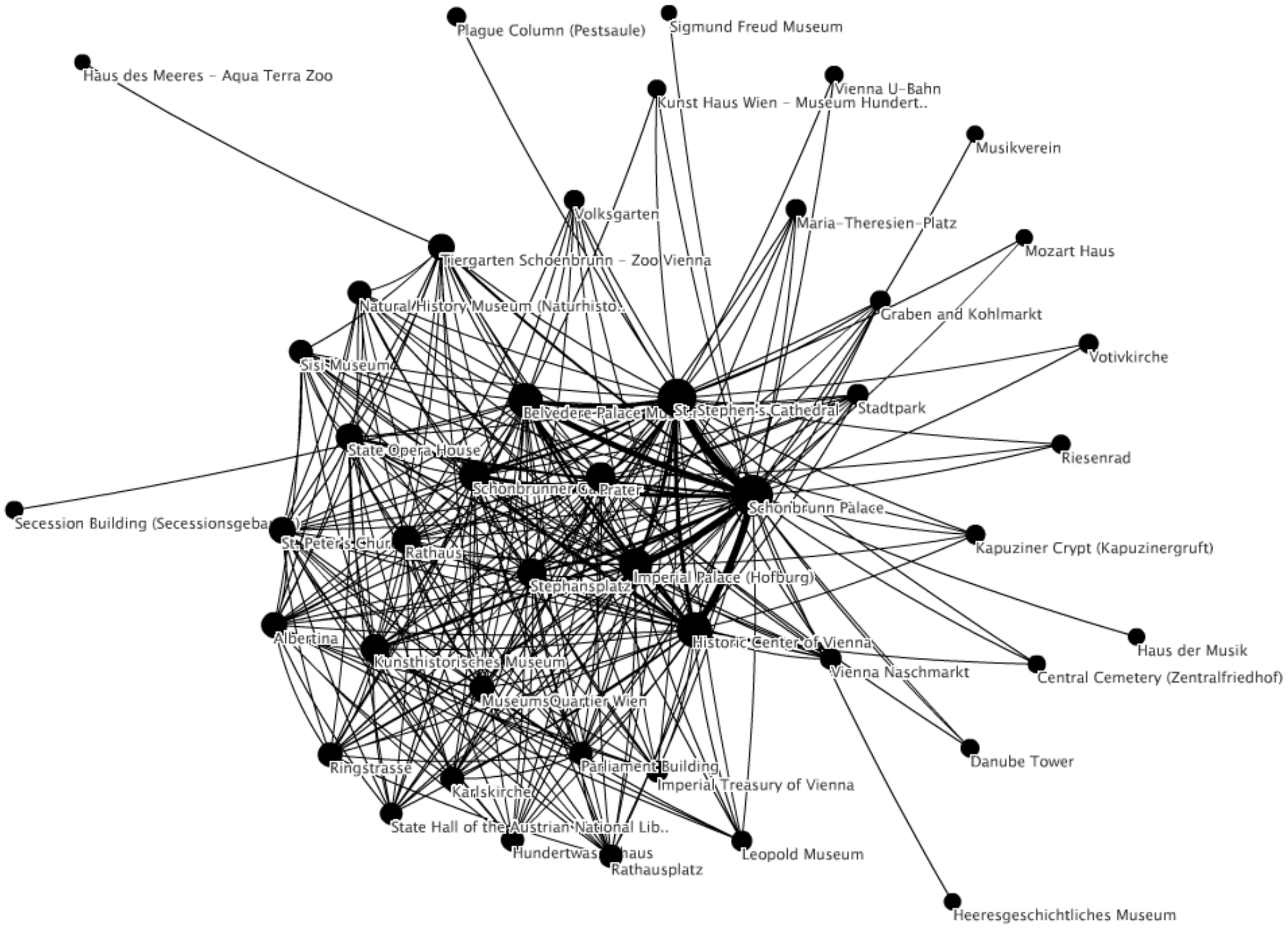

*Figure 6: The top 45 attractions and 300 co-occurrences of community A.*

To understand the results better we show the top 45 attractions in Fig. 6 on Vienna map in Fig. 7. Comparison of the two figures indicates that geographical distance between the attractions is not a crucial decision factor for visiting attractions within a destination. For instance, the attraction with the highest network flow, “Schonbrunn Palace”, is approximately 6 km away from the city centre, which is the location of the other three top attractions, namely “St. Stephen's Cathedral”, “Historic Center of Vienna” and “Imperial palace (Hofburg)”. Interestingly, the connections between “Schonbrunn Palace” and each of the other three attractions contain a larger volume of flow than the connections between the three attractions located in the city centre.

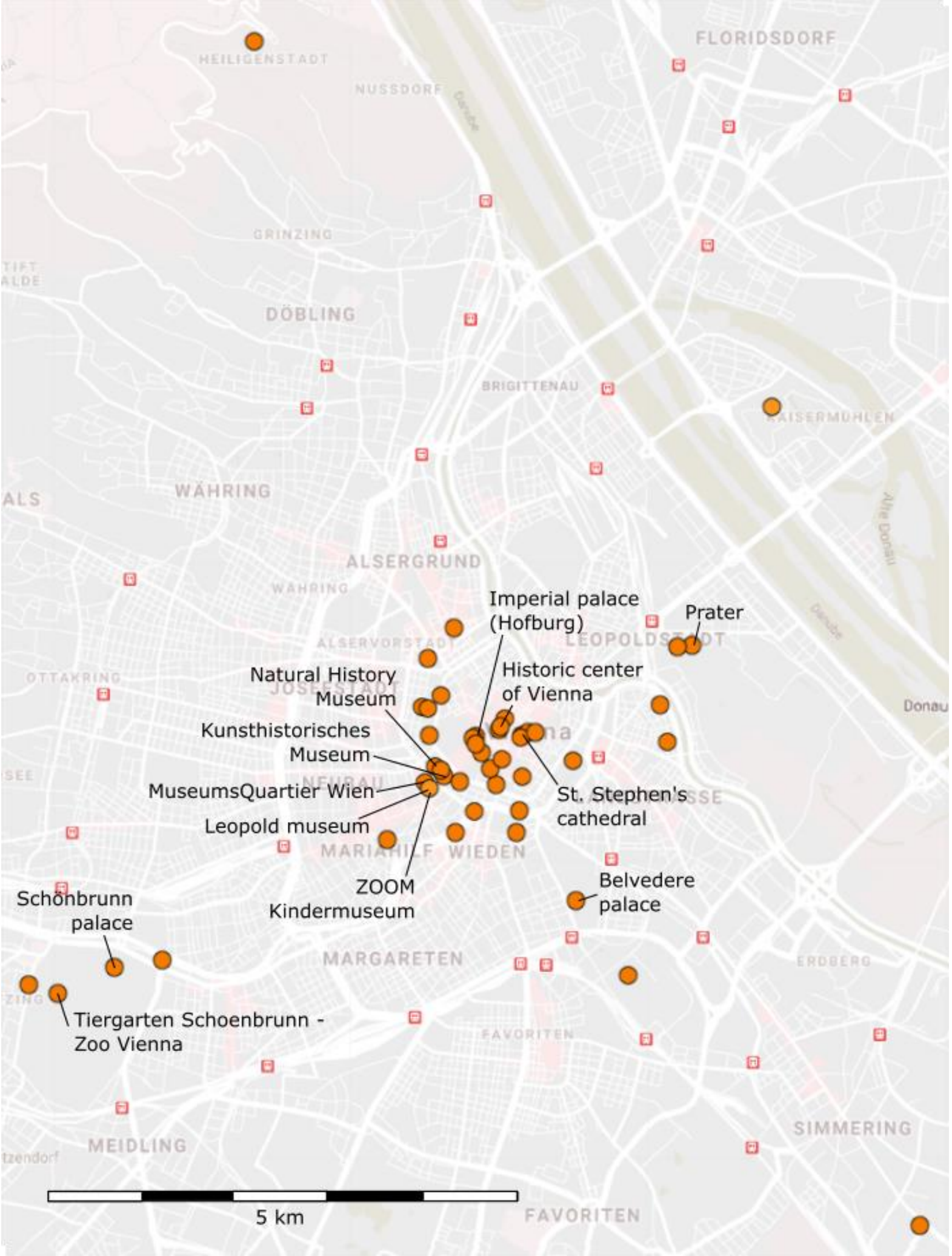

*Figure 7: Map showing the geolocations of the top 45 attractions from community A (shown with point symbol). Image: Google My Maps, Map data: ©2019 GeoBasis-DE/BKG (©2009), Google.*

The proposed approach also helps better understand the connections between attractions. To do this we select a certain attraction and observe its main conncetions to other attractions. We demonstrate this on the five museums, which are located in the city centre and in the vicinity of many major attractions (see Fig. 7). These are “MuseumsQuartier Wien”, “Kunsthistorisches Museum”, “Natural History Museum“, “Leopold museum” and “ZOOM Kinder Museum”. The five analysed museums are located within a 100-meter radius from each other. This is important as we wanted to minimize the influence of geographical distance on tourists’ decision to visit one of the five museums. For each of the five museums we analysed their top five connections by the flow volume.

The results are presented in Fig.8 where we can see five different attraction groups (marked as A1 – A5), each having one central attraction (the selected museum) and five connected attractions. The size of each attraction indicates the relative volume of the network flow through the attraction, while the thickness of each connection indicates the relative volume of the network flow between the two attractions with additionally shown percentage of the total network flow. The attractions are arranged around each museum in clockwise order based on the connections flow volume. The attraction with highest connection volume is shown on the top.

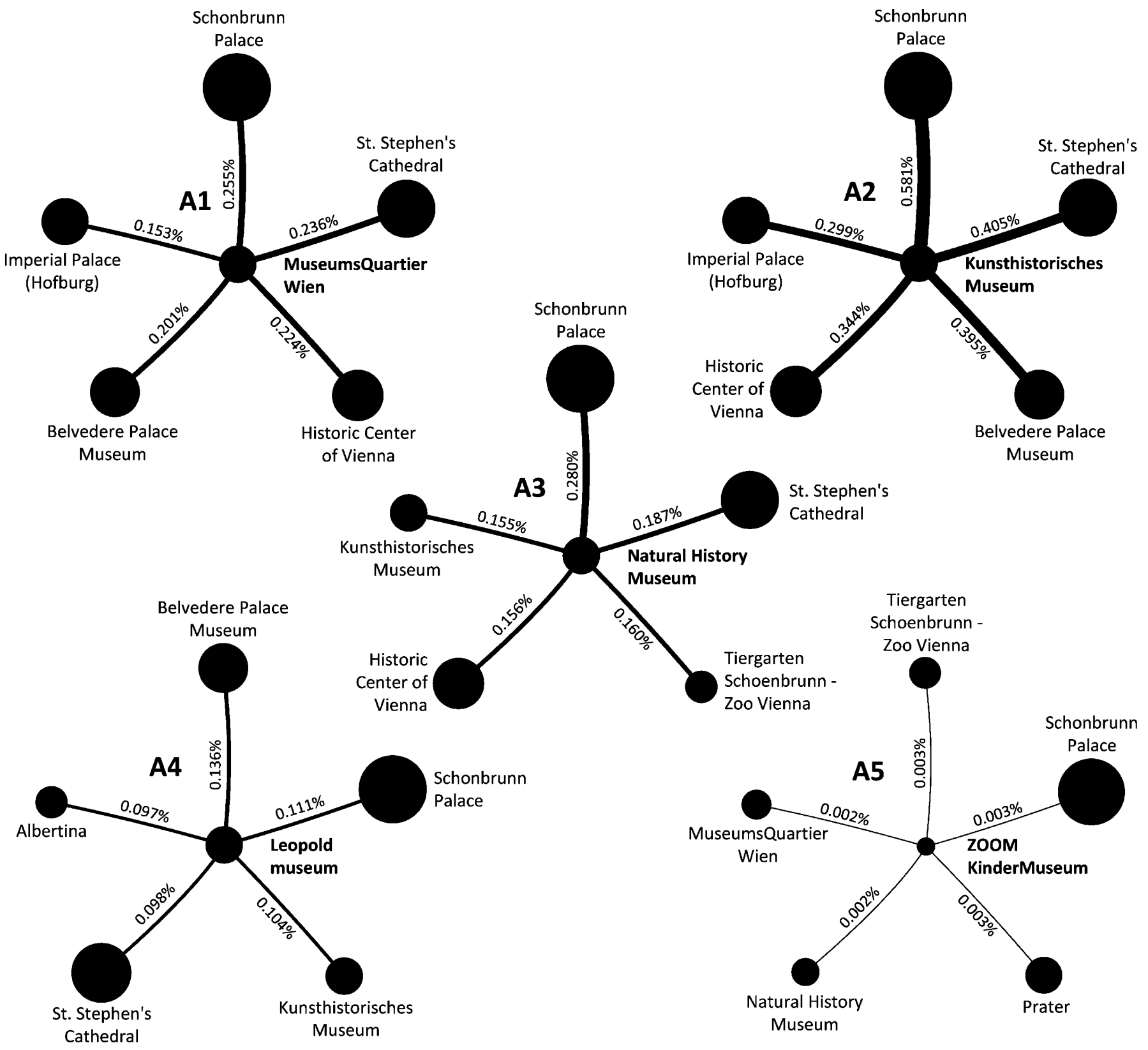

*Figure 8. The five attraction groups (A1-A5) for selected museums.*

Fig.8 shows that the order of the top five attractions connected to "MuseumsQuartier Wien" in group A1 is the same as the order of attractions by their network flow in Table 1. This is expected as the most visited attractions typically also have connections with high volume of network flow. For instance, we can expect "Schonbrunn Palace" to be the largest contributor to the visit of many other attractions. We can see a similar situation in the case of "Kunsthistorisches Museum" (A2) where the difference is that "Belvedere Palace Museum" and "Historic Center of Vienna" exchange places. Based on this analysis we see that "MuseumsQuartier Wien" (A1) and "Kunsthistorisches Museum" (A2) are interesting for tourists visiting main Vienna attractions. However, in case of the “Natural History Museum” (A3) “Tiergarten Schoenbrunn - Zoo Vienna” is the third most important attraction, although in Table 1 it is not among top 10 attractions in Vienna. This indicates a specific segment of visitors who are more interested in zoos and natural sciences. We can speculate that these probably include families and children.

As for “Leopold museum” (A4), the attraction with the highest contribution of network flow is no longer “Schonbrunn Palace”, but “Belvedere Palace Museum”. Additionally, “Kunsthistorisches Museum” takes the third place and “Albertina” the fifth. This clearly points to a different segment of visitors who enjoy contemporary art which is on display in “Belvedere Palace Museum”, “Leopold museum” and “Albertina”.

Finally, in the case of “ZOOM Kinder Museum” (A5) only “Schonbrunn Palace” remains one of the top five attractions from Table 1 that is part of the group. The largest part of the flow is generated by “Tiergarten Schoenbrunn - Zoo Vienna”, following “Prater” amusement park. This indicates different interests of the visitors of “ZOOM Kinder Museum”. We can speculate that these are typically families with young children who visit the main Vienna attractions less often.

Similar analysis can be performed for any attraction to discover the connected attractions that contribute the highest volume of flow.

## Limitations

This paper has several limitations. First, we only tested the proposed approach in the case of Vienna. Although in our opinion there is no theoretical obstacle that would limit the use of the approach on other destinations, further tests would be beneficial as they would help us obtain additional insights and maybe even general patterns of tourist movements typical for certain types of destinations.

Second, as we used data shared by visitors on TripAdvisor™, we only observed the visitations of the tourists that have posted on the tourism web platform. Although this can be considered as an important limitation of the approach, the existing research shows that even the tourists not posting about their experience are reading posted information (Bronner and de Hoog, 2016), which further affects their travel (Jin and Phua, 2016).

Third, data was collected over a relatively long period of time during which changes to TripAdvisor™ platform were made as for instance its user base moved from computers to smartphones. These changes could affect how and when tourist submit their posts. It is important to consider that the proposed approach assumes that all posts of the attractions visited during a single trip are submitted in a close time proximity to one another. Therefore, the approach works even if posts are submitted well after the trip, but in close time proximity to one another. However, this might not always be the case.

Fourth, there is always possibility of misleading and inaccurate posts, even though TripAdvisor™ monitors the posts. Tourists may also post only about certain attractions they visited, so the posts do not capture the full experience of tourist visits. We partially address this problem by removing poorly represented co-occurrences, as it unlikely that TripAdvisor™ would approve higher number of misleading posts about two particular attractions.

The last three limitations could undermine the relevance of the proposed approach and the results of this study. To address these limitations, we compared the top 20 attractions from the Table 1 to Vienna Tourist Board official ticket sales (Vienna Tourist

Board, 2008). The ticket sales data are available for the top 34 Vienna attractions in 2008. However, several important difficulties hinder this comparison. First, official data is available for paid attractions only, while no official statistics for freely accessible attractions is available. Second, many attractions are partially unpaid and sell tickets only to visit specific parts. For instance, general access to St. Stephen’s Cathedral is free, so official statistics only consider sales of tickets to its underground crypts and tower. Third, some attractions sell tickets valid for more than one entry. While in the official statistics some of them report the number of tickets sold, other report the number of visitor entries. Finally, so-called combination tickets are sold that are valid for several different attractions, but it is not clear if purchaser actually visited all of them. Considering these difficulties we categorized the attractions in Table 1 to three types: paid (Schonbrunn Palace, Belvedere Palace Museum, Kunsthistorisches Museum, Tiergarten Schoenbrunn, Albertina, Natural History Museum, Imperial Palace - Hofburg and Sisi Museum), partially unpaid (St. Stephen's Cathedral and Prater – Wiener Riesenrad) and unpaid (the remaining 10 attractions). While we were unable to compare the unpaid attractions, the analysis showed that the eight paid and the two partially unpaid attractions from Table 1 are also among the top 10 attractions according to the official ticket sales data. In our opinion, these results are a clear indication of the relevance of TripAdvisor™ data in the context of our study. Furthermore, the attractiveness of the top attractions is also related with the promotional activities. For instance, the eight paid attractions from Table 1 are all included in the official Vienna City Card and also among the “Must sees” in the official Vienna Tourist Board web page.

## Conclusion

The main contribution of this paper is a novel approach based on MBA to analyse user-generated data in tourism. The application of the proposed approach clearly showed that it can provide new and previously unavailable information important for tourism management and marketing. An advantage of the proposed approach is that it uses publicly available tourist posts with geo-locations to identify communities of attractions that are visited simultaneously. Moreover, the approach facilitates analysis of a specific attraction to identify the group of attractions, which contribute the highest volume of flow to this attraction. The approach is especially useful when studying tourism visitations in a limited geographic space with collocated attractions where the sequence of visits is mostly a consequence of casual tourist decisions, which hinders the use of the approaches that try to discover the dominant tourist movements.

Furthermore, our results clearly show that geographical location is not the crucial driver of visits. Attractions that are geographically close were not necessary part of the same attraction group. Groups were generated based on similar interests and not necessarily the distance between locations.

The largest community that our approach detected in Vienna contained over 99 % of the total network flow (see Fig. 4). This was at least partially expected as our analysis focused on a single city and it is in line with other studies showing that tourism is a highly concentrated industry. Further concentration tendencies can be observed at the level of attractions, where top attractions contain most of the flow (see Fig. 6).

In addition to testing the proposed approach on different destinations and using user-generated data from different web platforms, future work should focus on analysing other data that can be obtained from user posts, e.g. attraction evaluation, number of posts submitted by a user, user profile, other demographic data etc. Additionally, future work should consider seasonal patterns, weather patterns, day of week patterns etc. and especially linking these patterns to the economic impact. Such research, for instance, could: help improve tourists' experience by matching the tourists' profiles and existing visitations to the attractions that they are likely to enjoy depending on the season and the weather; help improve promotion of less known attractions while relieving overcrowded attractions again depending on the season and the weather; help decision makers to better understand destination tourism demand over season and plan supply accordingly; help decision makers to develop promotional campaigns that are focusing not only on main attractions of the destination; help decision makers to develop smart pricing strategies combining different attractions and developing attractive packages that create unique destination experiences etc.

Our study shows that applying MBA to user-generated data from tourism platforms provides relevant information on tourism demand patterns important for smart tourism decision-making. Using MBA we manage to understand which attractions are visited together the in city of Vienna. With ever-increasing quantity of data available from tourism platforms new avenues are opening in tourism research. Thus, researchers in tourism should focus not only on MBA, but also on other advanced analytical approaches that have been developed in recent years for the analysis of user-generated data and have been proven in other fields. In order to make smarter decisions in the future, better and more accurate analyses have to be conducted. Digitalisation and employment of existing and emerging technologies is providing us with tools that will improve our understanding of tourism of today and in the future.